\documentclass[conference]{IEEEtran}
\usepackage[utf8]{inputenc}
\usepackage{amsmath,amssymb,amsfonts}
\usepackage{algorithmic}
\usepackage{graphicx}
\usepackage{textcomp}
\usepackage{xcolor}
\usepackage{url}
\usepackage{authblk}
\def\BibTeX{{\rm B\kern-.05em{\sc i\kern-.025em b}\kern-.08em
    T\kern-.1667em\lower.7ex\hbox{E}\kern-.125emX}}

\newcommand{\GG}[1]{}

\begin{document}

\title{Deep embedded clustering algorithm for clustering PACS repositories}

\author[1,2]{Teo Manojlović}
\author[3,4]{Matija Milanič}
\author[1,2]{Ivan Štajduhar}
\affil[1]{\textit{University of Rijeka, Faculty of Engineering, Department of Computer Engineering}, Rijeka 51000, Croatia \authorcr Email: {\tt \{tmanojlovic, istajduh\}@riteh.hr}}
\affil[2]{\textit{University of Rijeka, Center for Artificial Intelligence and Cybersecurity}, Rijeka 51000, Croatia} 
\affil[3]{\textit{University of Ljubljana, Faculty of Mathematics and Physics}, Ljubljana 1000, Slovenia}
\affil[4]{\textit{Jozef Stefan Institute}, Ljubljana 1000, Slovenia}

\maketitle

\begin{abstract}
Creating large datasets of medical radiology images from several sources can be challenging because of the differences in the acquisition and storage standards. One possible way of controlling and/or assessing the image selection process is through medical image clustering. This, however, requires an efficient method for learning latent image representations.
In this paper, we tackle the problem of fully-unsupervised clustering of medical images using pixel data only. We test the performance of several contemporary approaches, built on top of a \textit{convolutional autoencoder} (CAE) -- \textit{convolutional deep embedded clustering} (CDEC) and \textit{convolutional improved deep embedded clustering} (CIDEC) -- and three approaches based on preset feature extraction -- \textit{histogram of oriented gradients} (HOG), \textit{local binary pattern} (LBP) and \textit{principal component analysis} (PCA). CDEC and CIDEC are end-to-end clustering solutions, involving simultaneous learning of latent representations and clustering assignments, whereas the remaining approaches rely on \textit{k-means} clustering from fixed embeddings.
We train the models on $30,000$ images, and test them using a separate test set consisting of $8,000$ images. We sampled the data from the PACS repository archive of the Clinical Hospital Centre Rijeka. For evaluation, we use \textit{silhouette score}, \textit{homogeneity score} and \textit{normalised mutual information} (NMI) on two target parameters, closely associated with commonly occurring DICOM tags -- \textit{Modality} and anatomical region (adjusted \textit{BodyPartExamined} tag).
CIDEC attains an NMI score of $0.473$ with respect to anatomical region, and CDEC attains an NMI score of $0.645$ with respect to the tag \textit{Modality} -- both outperforming other commonly used feature descriptors.
\end{abstract}


\section{Introduction}\label{s:intro}
Picture Archiving and Communication Systems (PACS) repositories contain vast quantities of historically recorded medical imaging data. The data stored in a PACS normally consist of medical images recorded on a patient using appropriate imaging techniques, and stored metadata information concerning the details on the conducted diagnostic procedures -- the latter being commonly stored using Digital Imaging and Communications in Medicine (DICOM) tags~\footnote{\url{https://www.dicomstandard.org/}}. Although DICOM tags facilitate the transfer of medical images between PACS repositories of different clinical centres, often they need to be supplemented with additional information and/or undergo manual intervention, for merging the data from different repositories~\cite{Dimitrovski2011}. This is influenced by the fact that DICOM tag values are subject to a significant variation due to common changes in clinical terminology, involving also the differences in diagnostic routines performed by specific clinicians/clinics, as well as the common language used.
Moreover, they can be incomplete, erroneous or missing. Therefore, to compensate, an additional source of information for merging data from multiple repositories is needed.

Content-based medical image retrieval (CBMIR) systems, which combine pixel data content with additional contextual information, can be used as a tool in medical decision support systems, as well as for mining large medical image datasets. In this paper, we explore several contemporary approaches for extracting and clustering latent image representations, comparing those that simultaneously optimise both the embedding (feature extraction) and the clustering assignments, with those that learn clustering assignments only after performing feature extraction.
We evaluate model performance on a sizeable dataset of medical images, using several evaluation metrics. The work presented here shows that useful latent image representations can be learned directly from pixel data using fully-unsupervised clustering techniques.

This work is structured as follows. In section~\ref{s:rel}, we describe related work. In section~\ref{s:m&m}, we describe the data, the modelling techniques and the evaluation metrics used in our experiments. In section~\ref{s:res}, we describe other aspects of the experimental setup and present and discuss the results obtained through qualitative, as well as quantitative evaluation. Finally, in section~\ref{s:con}, we conclude and propose future directions of research.

\section{Related work}\label{s:rel}
Because the number of stored medical images in clinical centres increases daily, organising them in a content-searchable format can be challenging, especially considering that they may differ in modality, anatomical region, patient position, as well as other characteristics. Clustering of images is one such method that can be used as the first step in analysing non-standardised and often mislabelled data, and it can be used as a starting point in building a CBMIR system.

When performing the clustering of images, there are two main steps that need to be taken, in order to achieve a satisfying clustering result. First, it is necessary to find a reasonably good feature descriptor. There are numerous papers investigating various feature descriptors. In~\cite{Veerashetty2019}, \textit{histogram of oriented gradients} (HOG) with \textit{Manhattan distance} is used as a feature descriptor for CBMIR of an MRI image dataset, attaining mean precision of $0.6$. In~\cite{Xu2009}, another commonly used feature descriptor -- \textit{local binary pattern} (LBP) -- is used for a similar purpose, attaining $92.5\%$ accuracy over a medical image dataset (specific details concerning the dataset used are omitted). In the last few years, deep learning has become a popular approach for learning feasible feature embeddings. In~\cite{Masci2011}, filters from a \textit{convolutional autoencoder} (CAE) are used to train a convolutional neural network, achieving a performance improvement on the CIFAR and MNIST datasets. In~\cite{Chen2017}, a CAE is used for learning features of CT images in an unsupervised fashion, which is later fine-tuned using a small amount of labelled data by training a classifier, attaining $95\%$ accuracy on a moderately-sized dataset of CT images.

Feature descriptors map input images from their original space to a latent space, where further processing can be performed. In~\cite{Zheng2004}, the authors propose an approach called locality preserving clustering (LPC). They show that it outperforms the \textit{k-means}-coupled \textit{principal component analysis} (PCA) model, attaining an NMI of $0.48$ on the test set consisting of general-purpose images (details are undisclosed), while \textit{k-means} attained an NMI of $0.43$. In~\cite{Ray2010}, \textit{k-means} and \textit{hierarchical clustering} are applied on a combination of pixel, local- and global-level features, attaining the highest accuracy ($92\%$) on clustering backbone X-ray images, and lowest accuracy on clustering hand X-ray images ($45\%$).

Recently, clustering techniques based on deep learning are becoming more popular. Here, feature learning and clustering is being done simultaneously, achieving state-of-the-art results on benchmark datasets. In~\cite{Xie2016}, \textit{deep embedded clustering} (DEC) algorithm is proposed, attaining $84.3\%$ accuracy on MNIST and outperforming \textit{k-means} coupled with an autoencoder. Focusing on some drawbacks of the DEC algorithm, \textit{improved DEC} (IDEC)~\cite{Guo2017b} is proposed, consisting of the original DEC loss coupled with the image reconstruction loss, attaining $88.06\%$ classification accuracy on MNIST. Later, the proposed model was further improved using the convolutional architecture~\cite{Guo2017a}. There also exist clustering methods based on other deep learning techniques, as well. E.g. in~\cite{Jiang2017}, a clustering method based on the \textit{variational autoencoder} (VAE) is proposed, attaining $94.46\%$ on MNIST and $84.45\%$ on STL-10 dataset. However, little or no work is reported concerning the application of fully-unsupervised deep learning techniques for clustering medical image repositories.



\section{Materials and methods}\label{s:m&m}
In this section, we present the main components of our work. In section~\ref{ss:data}, we describe the dataset used in our experiments. In section~\ref{ss:feat}, we describe the fixed feature extractors used for clustering. In section~\ref{ss:cae}, we describe the base CAE, used for learning and inferring base image embeddings for several clustering models. Next, in section~\ref{ss:clus}, we describe the (deep embedded) clustering approaches used in our experiments. Finally, in section~\ref{ss:eval}, we describe the evaluation metrics we used to assess model performance.

\subsection{Dataset and data preprocessing}\label{ss:data}
We use a sample of a larger dataset originating from the Clinical Hospital Centre (CHC) Rijeka, which consists of approximately $30$ million medical radiology images. From this dataset, we randomly sampled $30,000$ images to make the experiments more manageable. All images in the sampled dataset were additionally labelled using two related target parameters: (1) \textit{modality} (MOD) -- denoting the type of imaging modality used for image acquisition, directly taken from the DICOM tag \textit{Modality}, and (2) \textit{anatomical region} (AR) -- closely related to the DICOM tag \textit{BodyPartExamined}, supplemented by the information extracted from the \textit{StudyDescription} tag. These target parameters were chosen because they are simple, yet informative for differentiating between pairs of medical images. Moreover, confirming the assignment validity for both labels is fairly simple, even for a layman. Both target parameters are categorical variables. Other target parameters were not considered because their respective domains are much smaller, making them unsuitable for clustering evaluation. There were no missing values.


Because the sampled medical images vary in size greatly, we decided to resize them to standard dimensions $256\times 256$, while preserving their aspect ratios (zero paddings). Smaller-sized images lack detail, which makes them hard to compare visually; on the other hand, using larger images would have resulted in overly complex models and, therefore, would have required longer training times. We arrived at these conclusions through trial-and-error experiments on the validation subset. Pixel intensities were scaled linearly, per image, to the interval $[0,1]$.
We split the dataset into two disjoint subsets: a training subset consisting of $30,000$ images, and a test subset consisting of $8,000$ images. The validation subset is subsampled from the training subset, as required.

\subsection{Setting the latent embedding}\label{ss:feat}
One way of setting the latent image embedding is using fixed feature extraction techniques, such as PCA, HOG or LBP. In our work, we compare all three. Implementation details are described next.

PCA~\cite{Rehman2018,Hrzic2021} converts a set of possibly correlated variables into a set of linearly uncorrelated variables, i.e. principal components -- each component containing more variability than any subsequent component, in the provided data. Here, we used PCA to reduce the size of images to $20$ dimensions, consequently retaining at least $80\%$ of the explained variance on the training data.

HOG is commonly used for object detection in image processing~\cite{Dalal2005}. The descriptors are built using a non-linear function of image edge orientations in a dense grid, and then pooling into smaller, local-contrast normalised, spatial regions. The combined image histograms form the new representation. Here, we used $8 \times 8$ cells with $2 \times 2$ cells per block.

LBP is another commonly used feature descriptor in medical image retrieval systems~\cite{Xu2009}. This descriptor is built by dividing an image into cells where, in each cell, the observed pixel is compared to its neighbours within a set radius, which results in creating a histogram per image cell. Here, we used cells of size $16 \times 16$, radius was set to $1$ and the number of neighbouring points to compare was $8$.

\subsection{Learning the embedding function from data}\label{ss:cae}
\begin{figure}[!tb]
    \centering
    \includegraphics[width=1\columnwidth]{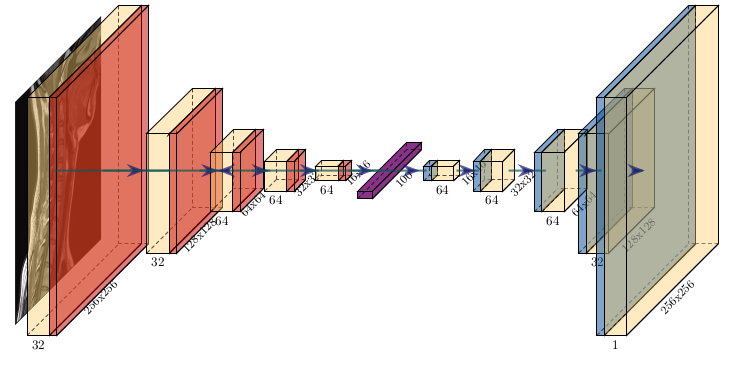}
    \caption{CAE model architecture used for learning the embedding function.} \label{fig:nn_architecture}
\end{figure}

As an alternative to predefined embeddings based on fixed functions, we trained a CAE~\cite{Masci2011,Chen2017} model as a common base for three clustering approaches described in section~\ref{ss:clus}. The model consists of the encoder part and a decoder part, connected to one another through a dense layer (Fig.~\ref{fig:nn_architecture}). The encoder part is the core feature extractor, mapping the images to a latent space, which is later fine-tuned using a specific clustering module for DEC and IDEC (section~\ref{ss:clus}). The module is trained on pixel data from preprocessed images that are part of the training subset (section~\ref{ss:data}), pursuing the goal of perfectly reconstructing the original images.

The encoder consists of five -- $3\times 3$ convolution, followed by a $2\times 2$ max-pooling -- layers. The first two layers have $32$ convolutional filters, each, while the remaining four layers have $64$ filters, each. The decoder has a symmetric, five-layer layout -- bilinear upsampling, followed by a convolution -- in each layer. The dense layer, connecting the decoder with the encoder, consists of $100$ neurons. We noticed that this model architecture attains satisfying reconstruction results while noticeably reducing the total number of model parameters. 
We use the \textit{ReLu} activation function in all layers, except the last; in the last layer, we use the \textit{sigmoid} activation function instead. Also, we made the following choices: we use \textit{mean-squared error} (MSE) as the loss function, \textit{Adam} as the optimiser, using a learning rate of $0.001$, and a mini-batch size $50$.
Model architecture and optimisation choices were determined by trial and error on the validation subset, using the values reported in related work for inspiration.



\subsection{Clustering latent representations}\label{ss:clus}
We compare the performance of the three algorithms based on traditional feature extraction, described in section~\ref{ss:feat} and coupled with \textit{k-means}, with several algorithms using a common CAE embedding (described in section~\ref{ss:cae}) -- which are described next. The algorithms using the common CAE embedding are the \textit{k-means}-coupled CAE, DEC and IDEC.

\textit{k-means} algorithm~\cite{Lloyd1982} iterates alternating improving clustering assignments and centroid values in the embedding space, until convergence. The embedding space is defined by the method used.

DEC algorithm~\cite{Xie2016} adapts the initial CAE embedding by minimising the \textit{Kullback–Leibler} divergence (KL) between the soft assignments $q$, and an auxiliary distribution $p$:

\begin{equation} \label{eq:kl_dec}
    L_n=\textrm{KL}(P||Q)=\sum_{i}^N\sum_{j}^K p_{ij}\log\frac{p_{ij}}{q_{ij}}, 
\end{equation}
where $N$ is the number of data points, and $K$ is the predefined number of clusters. Soft assignment $q_i$ is the similarity between the embedding $z_i$ and the cluster centre $\mu_j$, which is calculated using Student's t-distribution:

\begin{equation}\label{eq:dec_soft}
    q_{ij} =\frac{(1 + \left\Vert z_i - \mu_j \right\Vert^2)^{-1}}{\sum_{j'} (1 + \left\Vert z_i - \mu_{j'}\right\Vert^2)^{-1}},
\end{equation}
while the auxiliary distribution is calculated using the $p$ distribution:

\begin{equation} \label{eq:dec_aux}
    p_{ij} = \frac {q^2_{ij}/\sum_i{q_{ij}}}{\sum_{j'} {q_{ij'^2}/\sum_i{q_{ij'}}}}
\end{equation}

IDEC algorithm~\cite{Guo2017b} adds MSE to the original DEC loss (Eq. (\ref{eq:kl_dec})). In this equation, $x_i$ is the $i$-th datapoint, while $\hat{x}_i$ is the decoder output of the $i$-th datapoint:

\begin{equation}\label{eq:mse}
    L_r = \sum_{i=1}^n\left\lVert x_i - \hat{x}_i\right\rVert^2_2 .
\end{equation}

We minimise $L$ using stochastic gradient descent and back-propagation. We use $\alpha=0.1$ and $\beta=1$, as suggested in~\cite{Guo2017a,Guo2017b}.

To summarise, general expression defining the loss function of both DEC and IDEC is shown in Eq.(\ref{eq:cluster}), where the DEC loss can be be defined by setting $\beta=0$:

\begin{equation}\label{eq:cluster}
    L = \alpha L_n + \beta L_r, \quad \alpha,\beta \geq 0.
\end{equation}



By building DEC and IDEC on top of a pretrained CAE, we obtain the CDEC and CIDEC models. The training is finished when the maximum number of epochs is reached or the difference between losses in two consecutive epochs is less than $0.1\%$.

For all \textit{k-means}-based approaches (all models except CDEC and CIDEC), cluster assignments at inference are determined by assigning each point to the closest centroid using \textit{Euclidean} distance in the latent space, whereas CDEC and CIDEC use the posterior probability distributions to make equivalent assignments.

\subsection{Model evaluation metrics}\label{ss:eval}
We used several evaluation metrics to monitor the quality of clustering assignments. To monitor the cluster structure, we used \emph{silhouette score} (SS)~\cite{Rousseeuw1987}. SS measures how well the data are clustered, taking into consideration cluster tightness and the separation between clusters. It is calculated using the expression:
\begin{align}\label{eq:ss}
    SS(i) &= \frac{b(i) - a(i)}{\max\{a(i),b(i)\}}, \\
    a(i) &= \frac{1}{|C_i| - 1} \sum_{j \in C_i, i \neq j} d(i, j), \\
    b(i) &= \min_{k \neq i} \frac{1}{|C_k|} \sum_{j \in C_k} d(i, j).
\end{align}
Here, $a(i)$ denotes the mean distance between $i$-th instance and all other instances falling into the same cluster, and $b(i)$ denotes the smallest mean distance from $i$-th instance to all the instances not falling into the same cluster. $d(i,j)$ denotes the distance between a pair of instances $(i,j)$, and $|C_i|$ is the size of cluster $i$.

We used the \emph{normalised mutual information} (NMI) and the \emph{homogeneity score} (HS)~\cite{Rosenberg2007} to evaluate the semantic similarity of images falling inside specific clusters.
NMI is calculated using the expression:
\begin{equation} \label{eq:nmi}
    NMI(y, c) = \frac{I(y, c)}{\frac{1}{2}[H(y) + H(c)] },
\end{equation}
where $y$ represents ground truth labels, $c$ represents cluster labels, $I(y,c)$ represents the mutual information metric, and $H$ is the entropy. 
For showing the homogeneity of labels falling in specific clusters, we used HS, which falls in the range $[0,1]$: $1$ implying perfectly homogeneous clusters, and $0$ implying completely random cluster assignments. HS is calculated using the expression:
\begin{align}\label{eq:hs}
    HS(y, c)   &= 1 - \frac{H(C|K)}{H(C)}, \\
    H(C|K)      &= - \sum_{k=1}^{|K|}\sum_{c=1}^{|C|} \frac{a_{ck}}{N}log\frac{a_{ck}}{\sum_{c=1}^{|C|}a_{ck}}, \\
    H(C)        &= - \sum_{c=1}^{|C|} \frac{\sum_{k=1}^{|K|}a_{ck}}{n}log\frac{\sum_{k=1}^{|K|}a_{ck}}{n}.
\end{align}
Here, $K$ is the number of clusters, $C$ is the number of labels and $a_{ck}$ is the number of data instances belonging to the $k$-th cluster while being of class $c$.

To visualise specific embeddings in two-dimensional space, we use \textit{t-distributed stochastic neighbour embedding} (t-SNE) ~\cite{VanDerMaaten2008}, which is a non-linear alternative to PCA for visualising high-dimensional data in low-dimensional space. We consider specific embeddings with regard to the two categorical parameters, i.e. MOD and AR.

\section{Results and discussion}\label{s:res}
While searching for an adequate value of the number of clusters, we decided that $K=25$ is a reasonably good target number of clusters, with regard to the observed different-setup NMI scores on the validation dataset. Similarly, all hyperparameter values were determined through trial-and-error experiments on the validation data.

\begin{figure}[!tb]
    \centering
    \includegraphics[width=0.77\columnwidth]{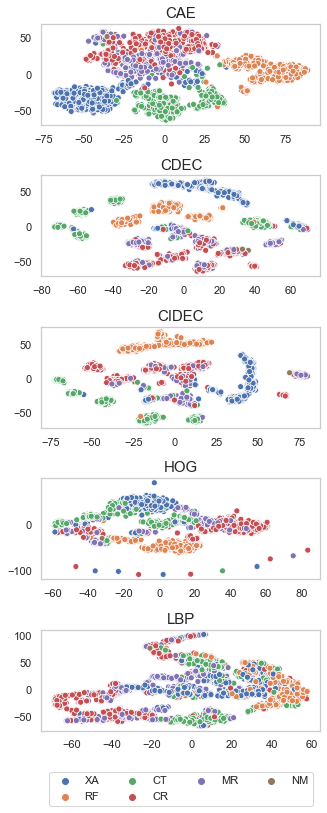}
    \caption{t-SNE visualisation of the embedding space with respect to parameter MOD for all approaches except PCA.} \label{fig:embMOD}
\end{figure}

\begin{figure}[!tb]
    \centering
    \includegraphics[width=0.9\columnwidth]{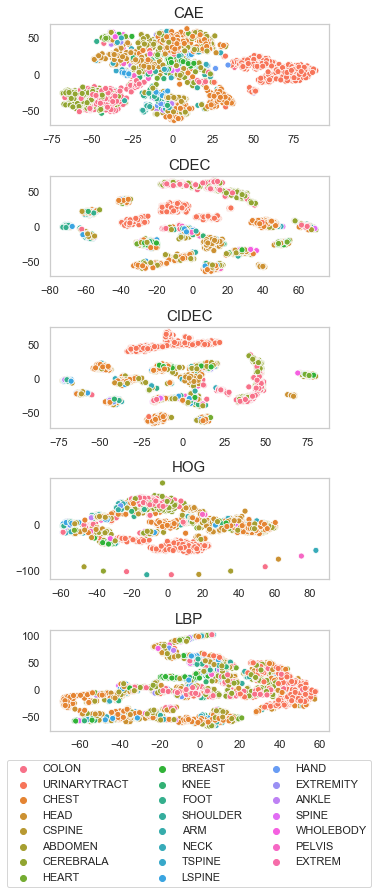}
    \caption{t-SNE visualisation of the embedding space with respect to the parameter AR for all approaches except PCA.} \label{fig:embAR}
\end{figure}

Qualitative characteristics of specific embeddings are observable in the t-SNE plots in Fig.~\ref{fig:embMOD} and Fig.~\ref{fig:embAR}. Both plots are generated based on models' performance on the test dataset.
In Fig.~\ref{fig:embMOD}, we can see that CDEC and CIDEC exhibit more homogeneous embedding and greater-sized clusters. On the other hand, HOG and LBP embeddings spread the space of specific clusters and seem to have trouble pulling a margin between specific clusters. CAE falls somewhere in between the two -- although it can be seen that the images belonging to the same MOD are close to one another, different modalities are not clearly separated. 
Similar observations hold for Fig.~\ref{fig:embAR} also. Here, we can see that clustering the images by AR is more demanding. However, it is much easier to extract useful information using CDEC or CIDEC because a clear margin between the majority of different clusters exists.

For the quantitative evaluation, to mitigate the effect of randomness induced during clustering initialisation on the evaluation metrics used, we trained each model in $10$ independent runs, and display the mean values for SS, NMI and HS in Table~\ref{res:comparison}. The results are calculated over the test dataset.
As expected, the models based on CAE perform better than the models based on traditional feature extraction mechanisms -- on both NMI and HS, for both target parameters.
Both evaluation metrics suggest that, although their performance is comparable, CIDEC performs best on clustering MOD, whereas CDEC performs best on clustering AR. It is also shown that both CDEC and CIDEC outperform all other models according to SS as well, producing cluster-oriented feature representations. Also, it can be seen that image clustering is more demanding for the target parameter AR, compared to the target parameter MOD.

\begin{table*}[!tb]
    \caption{Model performance with respect to NMI and HS for both target parameters (MOD and AR), as well as overall model SS. The results are calculated as means obtained from $10$ independent iteration runs. Best results are emphasised.}
    \centering
    \footnotesize
    \begin{tabular}{cccccc}
    \\
    \hline
    Algorithm & SS & NMI AR & HS AR & NMI MOD & HS MOD \\
    \hline
    CAE + k-means & 0.125 & 0.441 & 0.523 & 0.533 & 0.691 \\
    CDEC & \textbf{0.651} & 0.464 & 0.533 & \textbf{0.645} & \textbf{0.763}\\
    CIDEC & 0.622 &\textbf{0.473} & \textbf{0.544} & 0.637 & 0.755\\
    HOG + k-means & 0.146 & 0.394  & 0.451 & 0.526 & 0.659 \\
    LBP + k-means & 0.225 &  0.289 & 0.341 & 0.355 & 0.478 \\
    PCA + k-means & 0.201 & 0.351 & 0.342 & 0.482 & 0.482 \\
    \hline
    \end{tabular}
    \label{res:comparison}
\end{table*}

\section{Conclusion}\label{s:con}
In this paper, we explore the potential of fully-unsupervised clustering algorithms for clustering PACS repositories of medical images. We compare two state-of-the-art deep clustering methods built on top of a pre-trained CAE -- CDEC and CIDEC -- against CAE and other commonly used fixed feature descriptors, i.e. PCA, HOG and LBP, coupled with \textit{k-means}.
We conclude that CDEC and CIDEC outperform all other approaches on all used evaluation metrics and that they can be used as a starting step for creating large homogeneous datasets of medical radiology images.




Although the results presented in our study look promising, we believe that model performance of the deep learning approaches built on top of a pre-trained CAE can be further improved by using the information present in the DICOM tags associated with the images.


\section*{Acknowledgment}
This work has been supported in part by Croatian Science Foundation under the project \textit{IP-2020-02-3770}, by University of Rijeka under the project \emph{uniri-tehnic-18-15}, and by Republic of Croatia Ministry of Science and Education and Slovenian research agency bilateral project "Hyperspectral Image Analysis Using Machine Learning and Adaptive Data-Driven Filtering".

\bibliographystyle{IEEEtran}
\bibliography{export}

\end{document}